\documentclass[secnumarabic, graphics,floatfix, nofootinbib,tightenlines,nobibnotes, aps, pra, twocolumn,letterpaper]{revtex4}
\usepackage{graphicx}
\usepackage{epstopdf} 
\usepackage{amsmath}
\usepackage{amsfonts}
\usepackage{amssymb}
\usepackage[margin=0.5in]{geometry}
\usepackage[utf8]{inputenc}
\begin{document}

\title{Characterization of fluorescence collection optics integrated with a micro-fabricated surface electrode ion trap}

\author{Craig R. Clark}
\altaffiliation{Electronic mail: craclar@sandia.gov}
\author{Chin-wen Chou}
\author{A. R. Ellis}
\author{Jeff Hunker}
\author{Shanalyn A. Kemme}
\author{Peter Maunz}
\author{Boyan Tabakov}\altaffiliation{Center for Quantum Information and Control, University of New Mexico, MSC 07–4220, Albuquerque, NM 87131-0001}
\author{Chris Tigges}
\author{Daniel L. Stick}

\affiliation{Sandia National Laboratories, P.O. Box 5800 Albuquerque, NM 87185-1082}

\date{\today}

\begin{abstract}
One of the outstanding challenges for ion trap quantum information processing is to accurately detect the states of many ions in a scalable fashion. In the particular case of surface traps, geometric constraints make imaging perpendicular to the surface appealing for  light collection at multiple locations with minimal cross-talk.  In this report we describe an experiment integrating Diffractive Optic Elements (DOE's) with surface electrode traps, connected through in-vacuum multi-mode fibers.  The square DOE's reported here were all designed with solid angle collection efficiencies of 3.58\%; with all losses included a detection efficiency of 0.388\% (1.02\% excluding the PMT loss) was measured with a single Ca$^{+}$ ion.  The presence of the DOE had minimal effect on the stability of the ion, both in temporal variation of stray electric fields and in motional heating rates.  
\end{abstract}

\maketitle

\section{Introduction}\label{sec:Intro}
In recent years there have been multiple efforts to maximize the efficiency of fluorescence collection from single ions \cite{Almendros2009,Shu2010,Streed2011}.   Higher efficiencies lead to higher detection fidelities, shorter detection times, and increased rates of photon-mediated remote ion entanglement \cite{Maunz2009,Noek2013}. However, another aspect to consider is the ability to scale the optical system to simultaneously image multiple ions.  Single optics with large fields of view can image multiple ions but sacrifice collection efficiency and are not arbitrarily scalable.  Multiple optics each imaging separate locations are scalable if their individual lateral dimensions are comparable to the size of a single trapping well (the exact requirement depends on the specific architecture). In the case of micro-fabricated surface traps this requires the lateral dimensions of single optics to be smaller than $\approx$1 mm.  To simultaneously meet this requirement and still retain a high NA the lens must be commensurately  close to the ion ($<$1 mm).  However, this proximity is generally undesirable due to the physical constraints it imposes on the trap electrodes, the reduction of other optical access, and its impact on the trapping potential of the ion due to stray charge buildup. 

Most experiments use multi-element refractive lenses outside of the vacuum chamber for imaging ions.  These lenses are normally 10-30 mm away from the ion, collecting only a small fraction of the 4$\pi$ solid angle.  In previous experiments, a custom in-vacuum lens achieved a solid angle collection efficiency of 4$\%$  \cite{Almendros2009}, while a spherical mirror integrated with the trap resulted in a solid angle collection of up to 10$\%$ of the light from Ba$^{+}$ with a detection efficiency of 0.43$\%$ \cite{Shu2010}.  A micro-fabricated Fresnel optic (5 mm diameter, 3 mm working distance) with a solid angle collection of 12$\%$ (NA=0.64) resulted in an effective collection efficiency of 4.2$\%$ \cite{Streed2011}. Each of these methods maximize light collection for single ions in macro-scale traps. Other experiments have focused on demonstrating the size scalability of the collection optics.  A fiber integrated with a surface electrode trap yielded a solid angle collection of 2.1$\%$ \cite{VanDevender2010}, while a micro-mirror fabricated on a surface trap demonstrated a collection enhancement factor of 1.9 over their free space imaging system \cite{Merril2011}.  Our work integrated a fiber coupled lens array of micro-fabricated Diffractive Optic Elements (DOE's) with a 250 $\mu$m periodicity that is slightly smaller than the trapping well size, a proof-of-principle experiment combining scalable light collection with integrated micro-lenses.

In previous work we mechanically integrated a collection optic with a surface ion trap; reference \cite{Brady2011} details the optical design, integration with trap chip, various components of the vacuum chamber, and the approach taken to integrate all of the components into a complete system.  In this report we describe the lens characterization and improved alignment, as well as measurements of the stray electric fields and motional heating rates at different positions in the trap. These effects are significant concerns in the context of previous results using trapped ions as probes to measure the light-induced charging on dielectrics \cite{Harlander2010}.
	
\section{Experimental}\label{sec:Exp}
\subsection{Experimental Setup}
A two metal level technology was used to fabricate a surface electrode trap with two symmetric RF leads, 40 axial DC control electrodes, and two center DC control electrodes for principal axis rotation. More details on the trap fabrication and basic characterization can be found in \cite{Brady2011, Allcock2012}.  The trap is operated with a 250 V amplitude RF signal at a frequency of 35 MHz.  The DC voltages are supplied by six DAQ cards (National Instruments PXI 6733), as well as a controller (NI PXI 8106) and timing card (NI PXI 6652) to define and synchronize the waveforms for shuttling ions.  After compensation the resulting secular frequencies of the ion are [1.1, 5.4, 6.2] MHz.  

A neutral calcium flux is generated by resistively heating a tungsten wire wrapped around a ceramic tube containing natural calcium.  The source is mounted below the chip and the flux enters the trapping region through a 100 $\mu$m wide slot extending the length of the trap.  Calcium atoms are resonantly excited with a 423 nm diode laser followed by incoherent excitation to the continuum with a 375 nm laser.  Once trapped, the ions are Doppler cooled on the S$_{1/2}$-P$_{1/2}$ transition at 397 nm.  An 866 nm laser is used to re-pump from the metastable D$_{3/2}$ to the P$_{1/2}$ state (Figure \ref{fig:Pulse}(a)). The 397 nm laser is stabilized using a transfer cavity lock with Rb$^{87}$ as the atomic reference, and the 866 nm laser is software locked to a wavemeter which is calibrated every minute to the rubidium reference.

\begin{figure}[htp]
\includegraphics[width=\columnwidth]{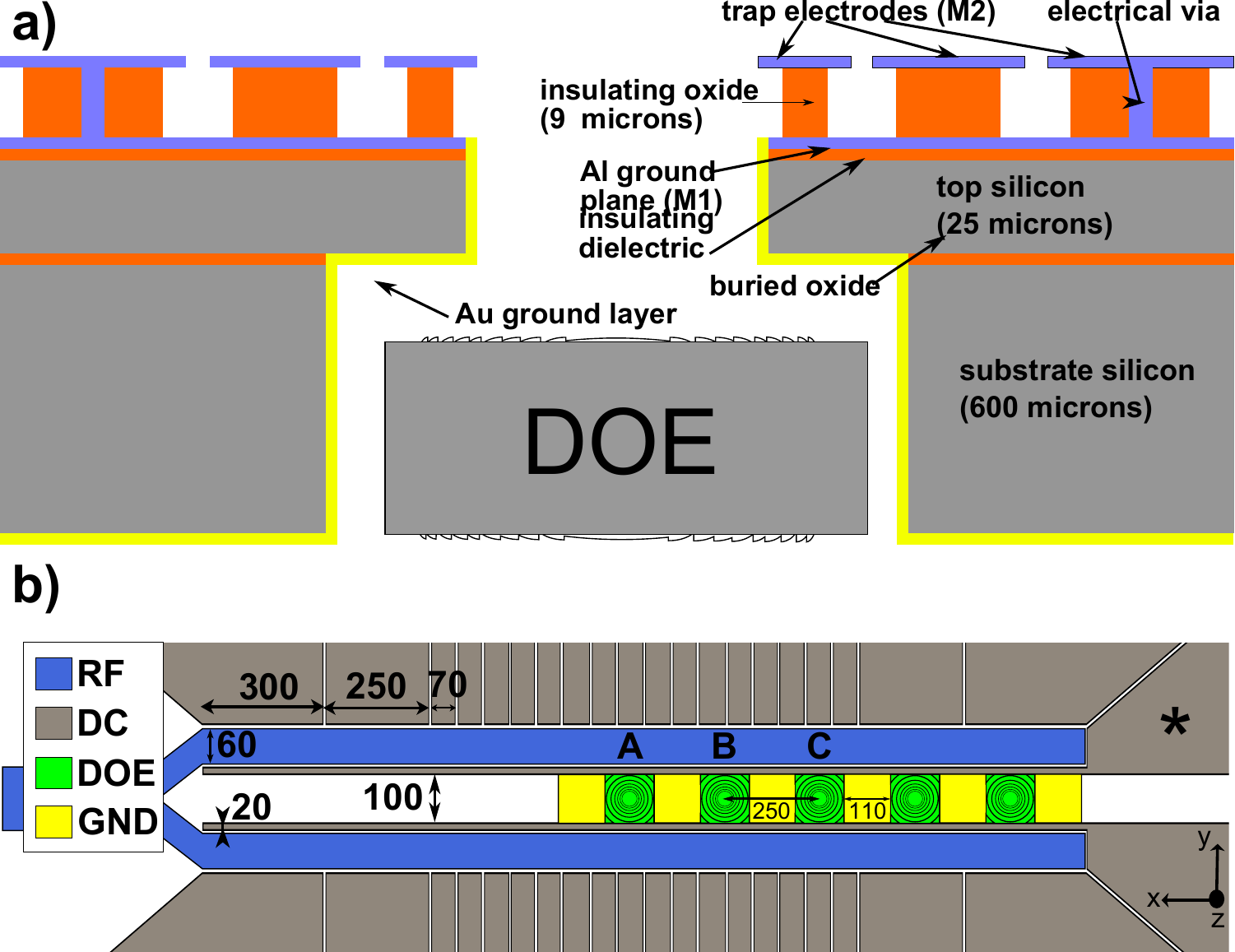}
\caption{\small(Color online) (a) Cross-sectional view of the ion trap (not to scale).  The ion is trapped 80 $\mu$m above the M2 surface with the DOE 85 $\mu$m below the M2 surface. (b) Plan view with dimensions shown in microns.  Green squares represent the DOE's and gold squares represent the gold ground planes between each DOE, which are electrically connected to the trap ground.} 
\label{fig:trap}
\end{figure}

	
A five lens array was mounted below the trap.  The stray field results described here were obtained using just the first three lenses and light collection was characterized with just the third lens.  The array is located 165 $\mu$m below the ion, and each 140 $\mu$m square lens is separated by a 110 $\mu$m gold ground plane from its neighboring lens, as shown in Figure \ref{fig:trap}.  The NA’s quoted in this report correspond to the stop in the multi-element system and assume a circular lens inscribed in the actual square lens.  However, the light collection percentages account for the square lens shape.  Each lens is aligned to a multi-mode fiber with a ceramic ferrule, which fixes the lens array's optical axis to the optical axis of the fiber. The fiber is connected to a UHV feedthrough for detection of the coupled light outside of the chamber.

\subsection{Experimental Characterization} \label{sec:EXCH}
We were primarily interested in characterizing three properties of the trap-optic system: the difference in stray electric fields above and away from the exposed dielectric of the DOE, the motional heating rate difference above and away from the DOE, and the light collection properties of the DOE.   

Excess micromotion occurs when the ion is not positioned at the RF null, due to a phase mismatch of the RF on different electrodes, an imperfect static trapping solution, or a stray electric field.  The first issue can be practically eliminated by capacitively shunting the static control electrodes, while the latter two can be eliminated with a static offset applied to the electrodes, provided the stray field does not change on a fast time scale.  The stray electric fields are compensated by ensuring the ion is positioned at the RF null at multiple locations along the linear trap using an adaptive algorithm with iterative measurements of the micromotion along the two transverse axes.  First the ion's position is coarsely compensated by scanning the frequency of the 397 nm laser and minimizing the micromotion sidebands of the atomic resonance.  Then a time-of-arrival technique is used to fine tune the compensation; the combination of these two methods allows us to compensate stray fields in the plane parallel to the trap (x,y) \cite{Berkeland1998}.  In the z-direction, the electric field is compensated by minimizing the motional excitation when applying a voltage resonant with the radial secular frequency to the RF electrode \cite{Ibaraki2011,Narayanan2011}.  Combining these techniques allows us to compensate stray fields down to the order of a few V/m.  After this procedure the applied voltage solution is compared to boundary element simulations of the trap to estimate the stray field in both y and z directions.      

A well known issue with ion traps is the motional heating due to electric field fluctuations on the electrodes \cite{Hite2012}.  This is particularly problematic in traps that confine ions close to trap electrodes (80 $\mu$m in the current work). The experiments described here employ a Doppler re-cooling technique \cite{Epstein2007} to measure and compare heating rates at different locations. 

\begin{figure}[htp]
\includegraphics[width=\columnwidth]{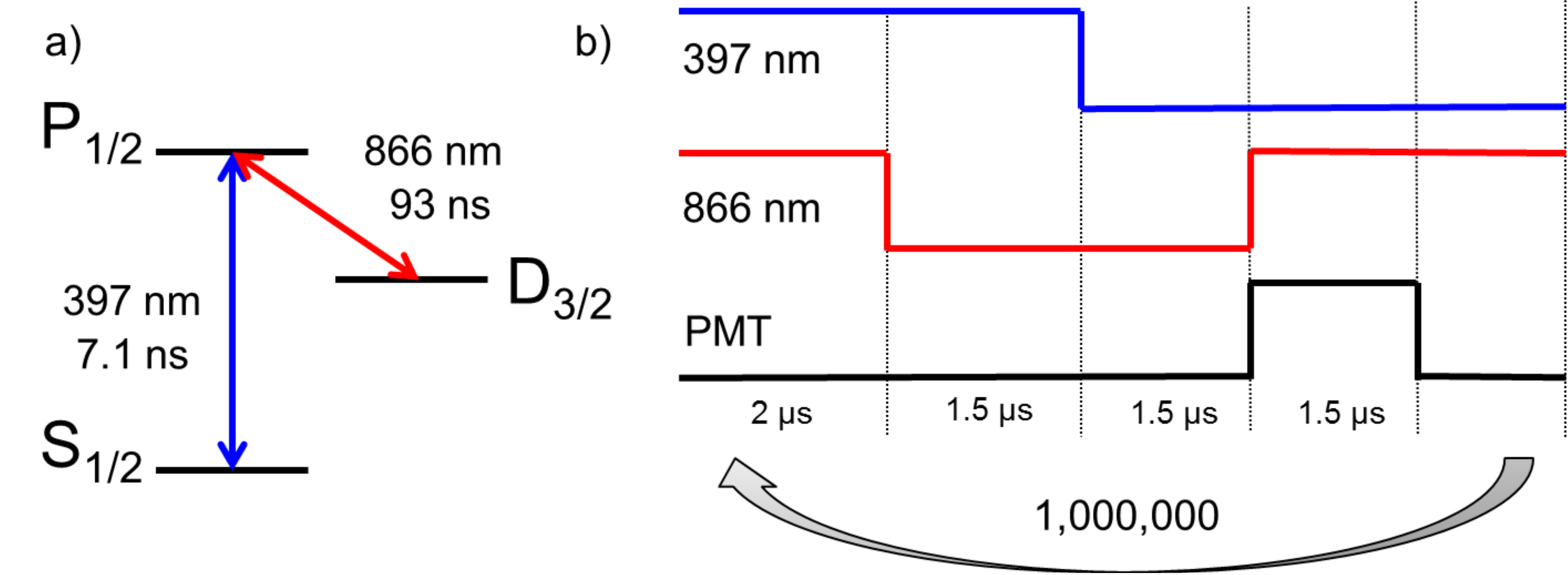}
\caption{ \small(Color online) (a) Ca$^{+}$ energy diagram showing only the relevant Doppler cooling transition and those lasers used in the paper. (b) Single photon generation pulse sequence. The ion is cooled for 2 $\mu$s with both lasers. Then it is exposed to a 397 nm laser for 1.5 $\mu$s to optically pump to the D$_{3/2}$ state. A 1.5 $\mu$s delay is inserted to ensure the previous laser is fully extinguished. The 866 nm laser is turned on to pump to the P$_{1/2}$ state, generating a single 397 nm photon which is detected by a PMT.} 
\label{fig:Pulse}
\end{figure}

To determine the collection efficiency of the optic a single photon counting technique was used \cite{Almendros2009,Shu2010}.  This technique does not require fitting an atomic spectra to an eight level optical Bloch equation.  The ion is prepared in the the metastable D$_{3/2}$ state and then re-pump to the P$_{1/2 }$ state which decays and emits a single photon (Fig. \ref{fig:Pulse}(a)).  Figure \ref{fig:Pulse}(b) shows a schematic of the pulse sequence along with the corresponding times associated with each step. This sequence is repeated $10^6$ times for statistics.
	
\begin{figure}[htp]
\includegraphics[width=\columnwidth]{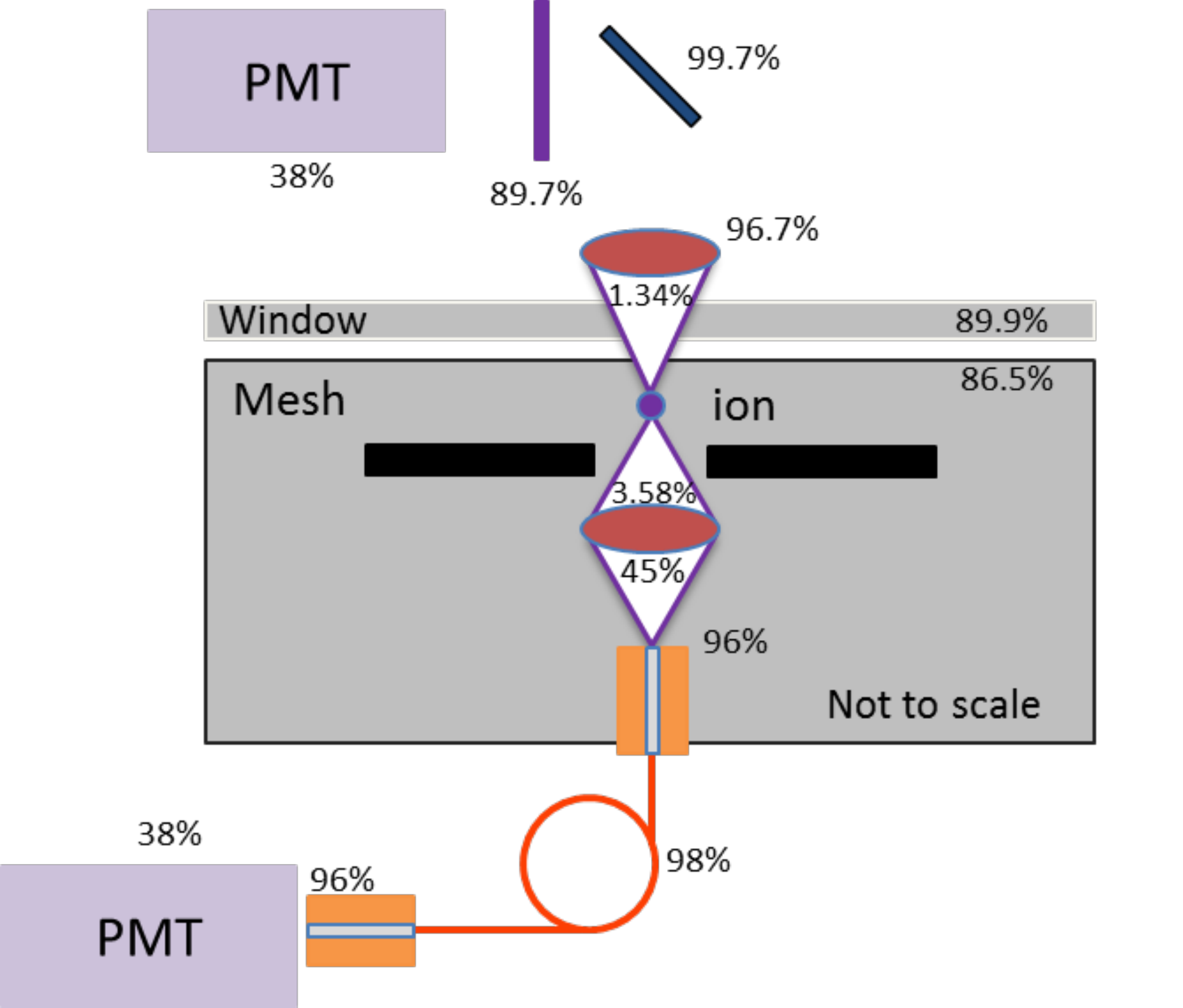}
\caption{\small(Color online) Schematic of the standard imaging system and DOE collection system (not to scale).  All percentages correspond to the transmittance or reflectivity at 397 nm.  The product of each stage results in a total detection efficiency of a trapped fluorescing Ca$^{+}$ ion.} 
\label{fig:OpticalSet}
\end{figure}

Figure \ref{fig:OpticalSet} is a schematic of both the standard imaging system and the DOE collection set-up, along with the transmittance of each optical component. Our standard imaging system includes an 86.5\% transmissive grounded mesh to reduce charge build up on the re-entrant view port.  The view port on the vacuum chamber is 89.9\% transmissive at 397 nm.  Our imaging optic is a UV objective from Special Optics (54-17-29-397) with NA = 0.29 (effective solid angle collection of 1.9\%) and 96.7\% transmittance). When combining the viewport and optic the effective solid angle collection is 1.34\%.  The collected fluorescence reflects off a UV coated mirror (99.7\% reflectivity) and passes through a Semrock 395/11 nm filter which is 89.7\% transmissive at 397 nm.  Using a flip mirror, all collected light is either be sent to an Andor Luca-R CCD camera for imaging the ion or a photon counting PMT (Hamamastu H10682-210, QE=38\% at 397 nm).  Considering all losses an end-to-end (ion-to-PMT) detection efficiency of 0.34\% of the light is expected.        

\begin{figure}[htp]
\includegraphics[width=\columnwidth]{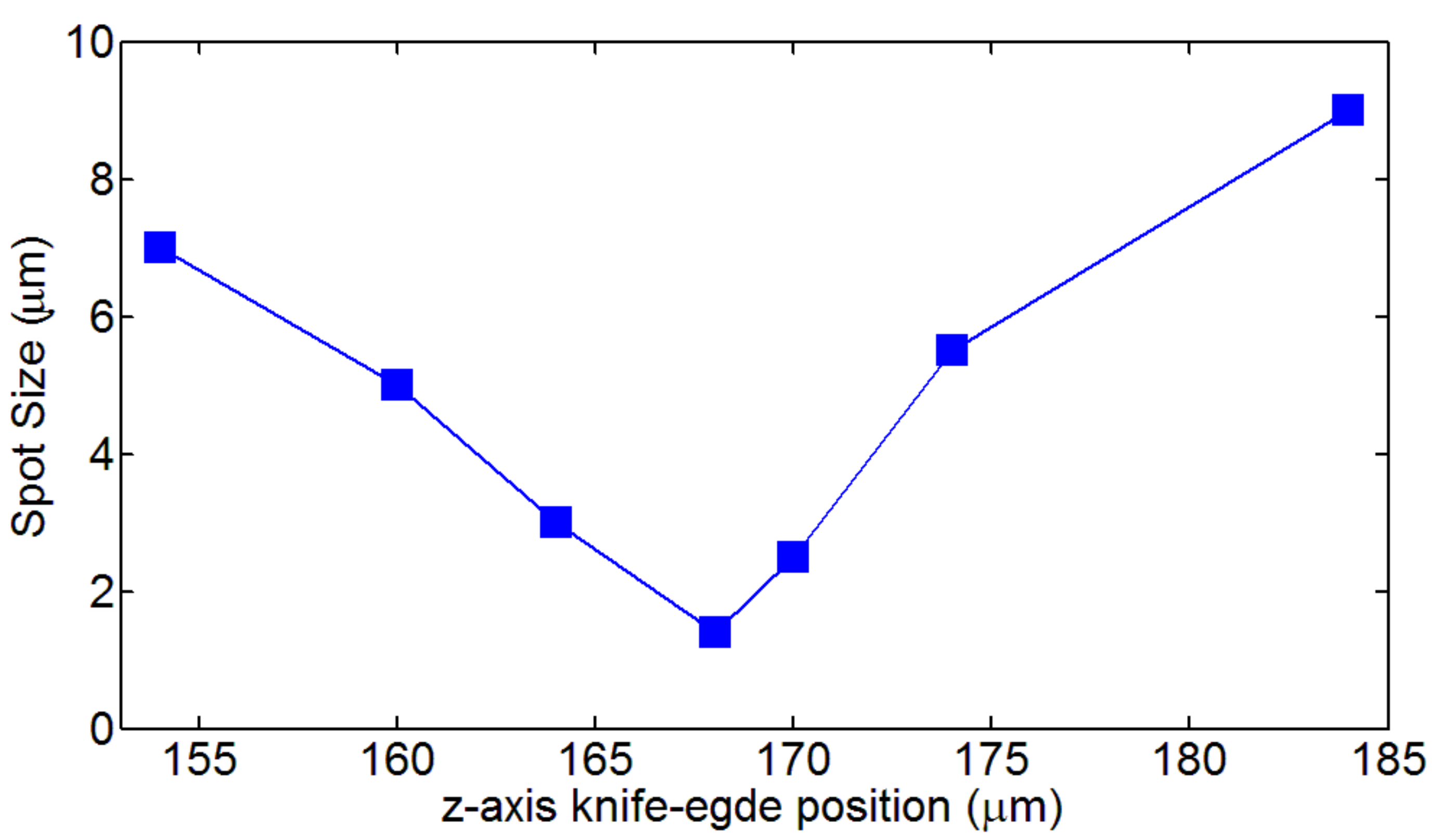}
\caption{\small(Color online) Spot size (diameter) as a function of distance from the collection DOE, with a line to guide the eye. A series of knife edge measurements were made at different distances from the collection DOE to determine the effective focal length.} 
\label{fig:Rayleigh}
\end{figure}

In designing the DOE lens system the most important metrics are the solid angle collection efficiency and diffraction efficiency.  The solid angle collection efficiency of the lens system reported here is 3.58$\%$. A standard knife-edge test was used to measure the diffraction efficiency, i.e. the fraction of light transmitted by the lens into the spot as a function of the position of the knife-edge at the focal point and various distances from the DOE (see figure \ref{fig:Rayleigh}).   The range of diffraction efficiencies of 15 fabricated and measured two-lens elements (in series) was 44-49\%, while the integrated lens had an efficiency of 45\%. The DOE was designed to have a focal distance of 164 $\mu$m  but was measured to be 168 $\mu$m. The collected light is coupled to a multi-mode optical fiber  (Polymicro’s FVP100110125 fiber: 100 $\mu$m core, NA = 0.22) in which 4\% loss is expected on the input and output facets and approximately 2\% absorptive loss through the fiber. Using the same PMT as above an end-to-end detection efficiency of 0.55\% is predicted.

\section{Results and Discussion}\label{sec:RandD}

\subsection{Stray field measurements and compensation}
UV light in proximity to exposed dielectric surfaces can lead to charge build-up which is dissipated slowly relative to the experimental time.  Oxides or contaminants on the trap electrodes can also become charged and generate electric fields at the ion, leading to excess micromotion which negatively impacts Doppler cooling, light collection, and the motional heating rate of the ion.  Although the ion has a direct line of sight to the dielectric, it is also significantly shielded by the surrounding electrodes and the grounded gold coating in the space between each DOE, as shown in Figure \ref{fig:trap}.   
	
\begin{figure}[htp]
\includegraphics[width=\columnwidth]{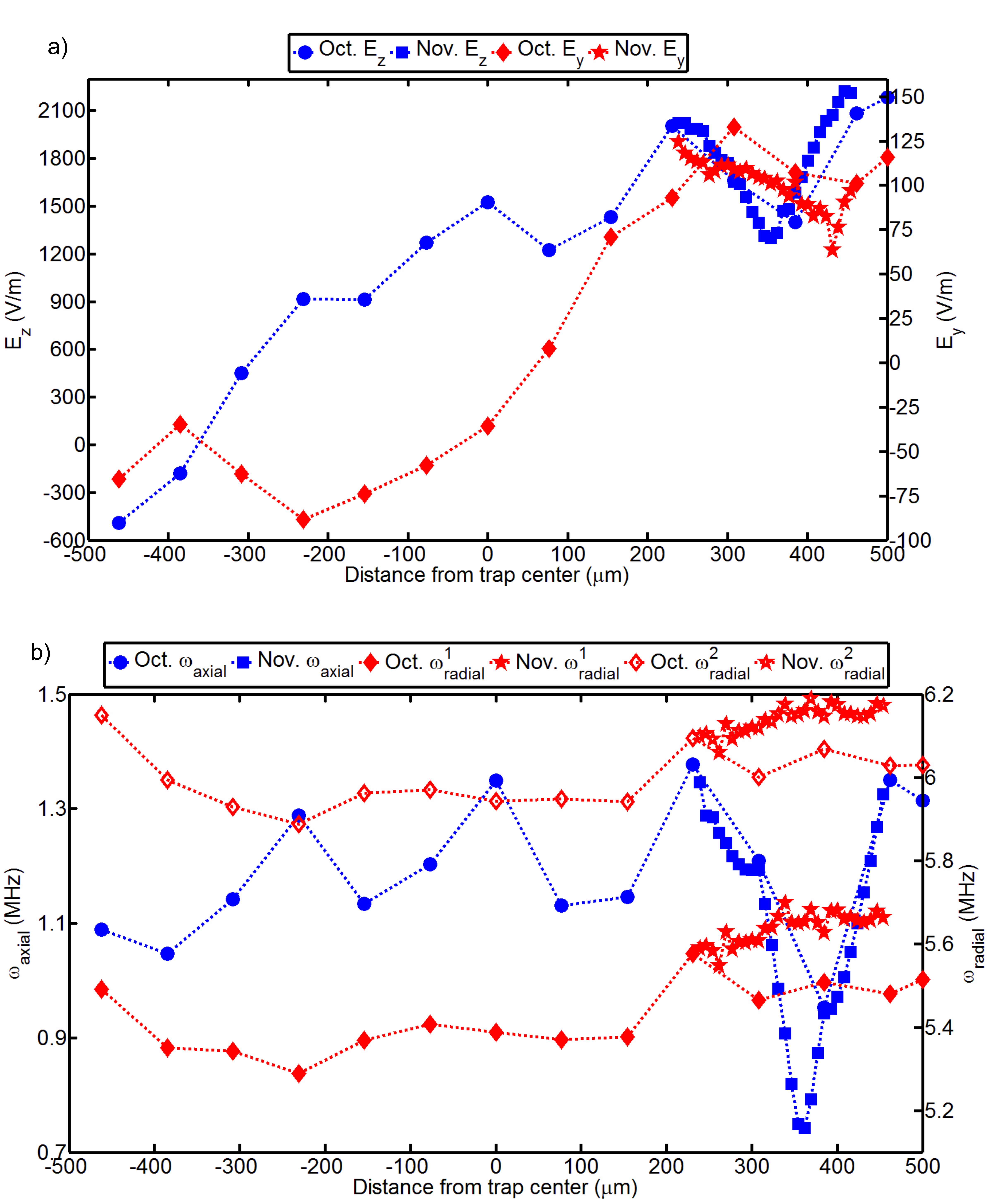}
\centering
\caption{\small(Color online) (a) Stray field compensation. The stray field is estimated by comparing our experimental results to our simulation results, with a dotted line to guide the eye.  E$_{z}$  (blue) and E$_{y}$ (red) as measured in the the trap.  (b)Secular frequency data taken along the z-axis of the trap after performing the compensation procedure.  $\omega_{axial}$ (blue) and $\omega_{radial}$ (red) are shown with a dotted line to guide the eye. Coarse data was taken at 77 $\mu$m steps (blue circles and red diamonds), higher resolution data (blue squares and red stars) was taken between the center of two DOE's at a step size of 5 $\mu$m.  The higher resolution data was taken a month after the coarse data.} 
\label{fig:Stray}
\end{figure}

Figure \ref{fig:Stray} shows measured stray fields in the $y$ and $z$ directions (radial to the RF confining potential) after performing an automated compensation procedure, compared away from and above the DOE's.  Stray field measurements were taken at a coarse spacing of 77 $\mu$m over a total range of $\pm$500 $\mu$m from the center of the $x$ axis of the trap, along with higher resolution measurements at a 5 $\mu$m spacing between optics $B$ and $C$.  
	
In Figure \ref{fig:Stray}(a), there is a steady increase in the applied field needed to position the ion at the RF null in both the $y$ and $z$ directions, corresponding to the ion's position relative to the DOE's. This is primarily due to a modification of the trapping potential by the DOE assembly, as compared to above the slot at -300 $\mu$m.  When measured above the slot the stray field $E_{z}$ is the same order of magnitude as $E_{y}$, but as the ion moves above the DOE assembly, $E_{z}$ dramatically  increases.  At positions 0, 250, and 500 $\mu$m there are local maxima in $ E_{z}$ of 1500, 2000, and 2200 V/m, which correspond to trapping positions directly above the center of optics $A$, $B$, and $C$, respectively. The local minimum of $E_{z}$ of 1300 V/m at 350 $\mu$m corresponds to storing the ion directly above the gold coated ground plane between $B$ and $C$.  The secular frequencies show a similar trend due to the stray field. The axial frequency has a local maximum and local minimum corresponding to positions above the DOE and ground plane, respectively. Figure \ref{fig:Stray}(b) also shows the radial frequencies at each location. The increase in radial frequency shown on the high resolution data is ascribed to changed in the helical resonator over the month period in which data was taken.      


To determine the impact of the DOE's on the temporal stability of the stray electric field, the automated compensation procedure was continuously performed over several hours at locations away from (-385 $\mu$m) and above DOE-$C$ (500 $\mu$m). The stray field was calculated by comparing the experimentally determined DC voltage solution which minimized micromotion with the simulated voltage solution.   There was little change in $E_{z}$ and $E_{y}$ (except for after loading an ion) over the course of five hours.  The changes due to loading resulted in $\Delta E_{z} \approx \pm 25$ V/m and $\Delta E_{y} \approx \pm 15$ V/m at both locations.  The axial frequency changed less than 2 kHz and the radial frequency changed less than 50 kHz over the course of five hours, including changes due to loading. The changes in the radial frequency are ascribed to instabilities in the resonance frequency of the RF resonator.   


\subsection{Motional heating}

\begin{figure}[htp]
\includegraphics[width=\columnwidth]{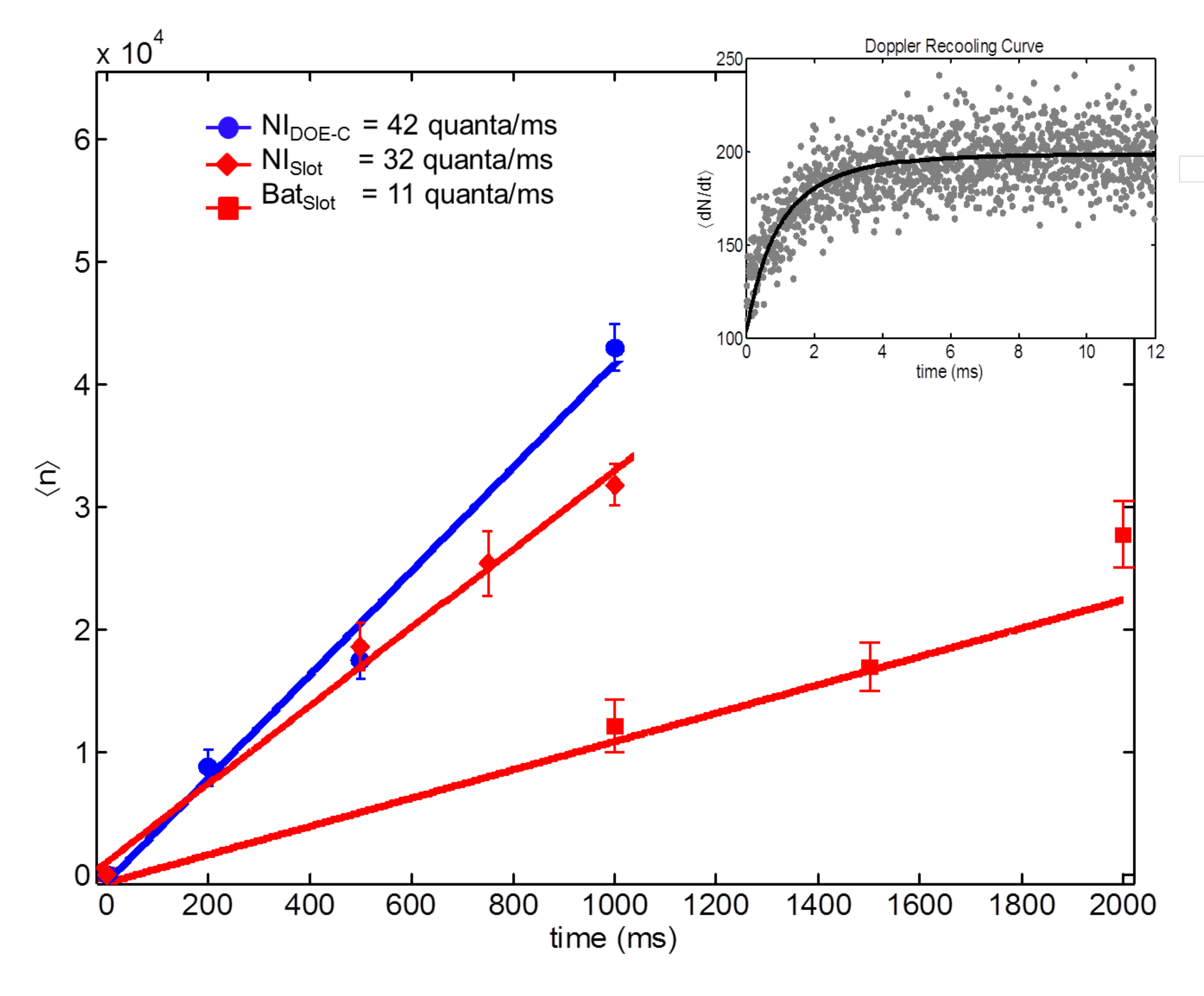}
\caption{\small(Color online)  Typical heating rate results (circles) when using the NI chassis to confine ions above DOE-$C$ (blue) and in the loading slot (red).   A heating rate of 11 quanta/ms was measured using a battery voltage supply (squares) to trap in the loading slot.  (inset) Doppler re-cooling curve with the ion positioned above DOE-$C$, with a heating time of 500 ms.}  
\label{fig:HeatingRate}
\end{figure}

Figure \ref{fig:HeatingRate} shows heating rate measurements in the slotted region of the trap and above DOE-$C$.  The measurements performed using the NI voltage sources were similar above the slot and optic (32 and 42 quanta/ms, respectively).  A measurement with a battery voltage supply in the slotted region yielded a heating rate of 11 quanta/ms.  The battery supply does not allow shuttling the ion so results above DOE-$C$ were not measured with the battery. 

\subsection{Light detection measurements}
\begin{figure}[htp]
\includegraphics[width=\columnwidth]{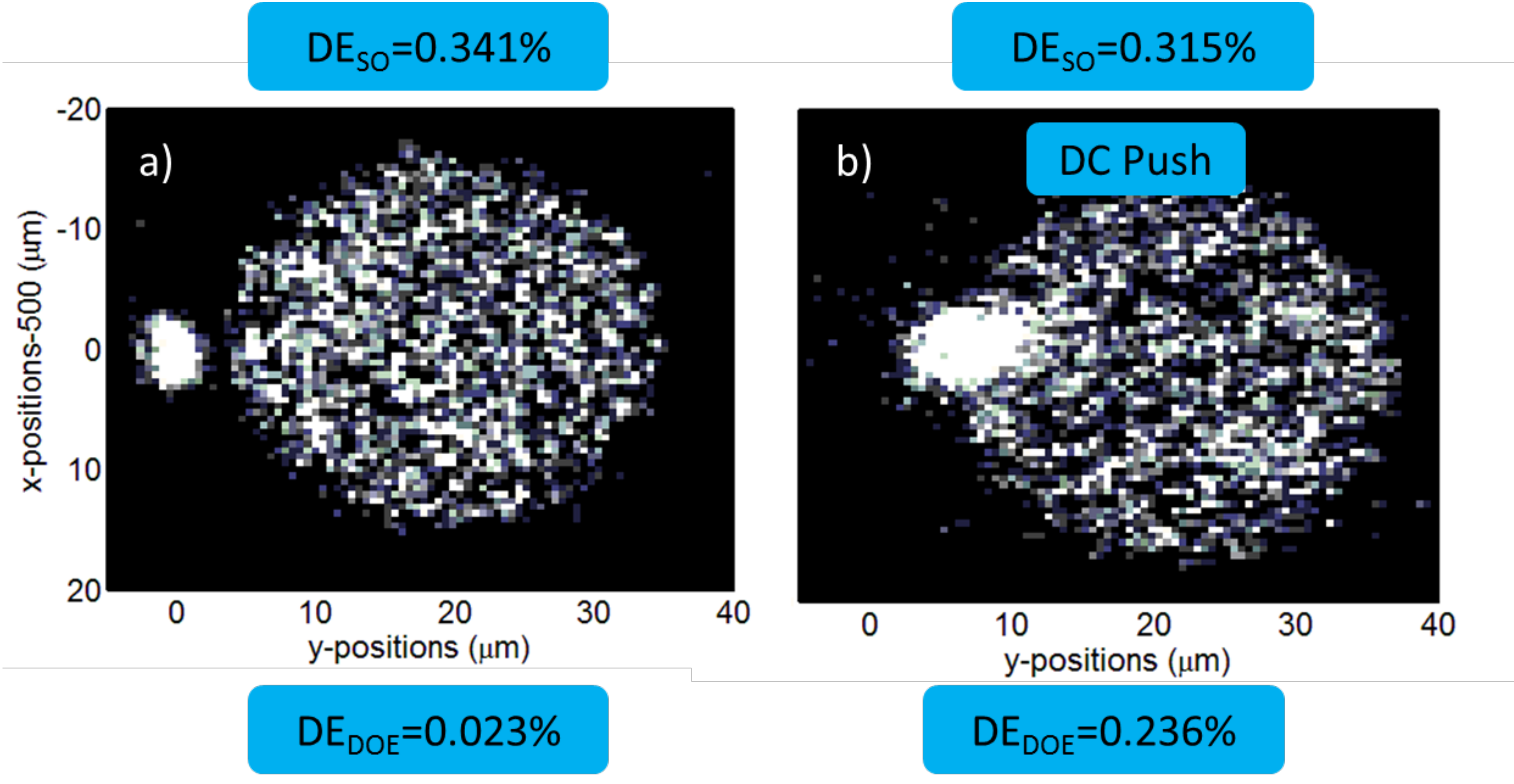}
\caption{\small(Color online) (a) Image of the compensated ion with corresponding $DE_{SO}$ and $DE_{DOE}$ (b) Image of an ion after applying voltage to the DC electrodes to shift the ion off the RF null by 7 $\mu$m (the smearing of the ion is due to the micromotion). The large circles in both cases correspond to 397 nm light which is back illuminating the optic through the fiber.}
\label{fig:DCImage}
\end{figure}

The light detection efficiency of the standard optics system was measured to be $DE_{SO}$ = 0.341(6)\%, which matches the theoretical value calculated from the product of the percentages in Figure \ref{fig:OpticalSet}.  The detection efficiencies reported here include all system losses, from the solid angle collection to the quantum efficiency of the PMT. Figure \ref{fig:DCImage}(a) shows an image of the ion after stray electric fields have been cancelled, along with 397 nm light back illuminating the DOE through the multi-mode fiber.  It shows that the compensated ion is not at the focus of the DOE, but is translated in the $y$ direction by almost 20 $\mu$m.  This was confirmed after removing the trap from the vacuum chamber and measuring an 18(2) $\mu$m translation, which likely occurred during the bake of the chamber.   The height of the focus was in-situ measured to be at the height of the ion.  It should be noted that the back illumination size is larger than the focal spot size shown in figure \ref{fig:Rayleigh} due to the size of the multi-mode fiber.  For more details on the pre-installation alignment procedure see \cite{Brady2011}. 
		
Additional voltage was applied to the DC control electrodes to shift the ion towards the optic by $\approx$7 $\mu$m, as shown in Figure \ref{fig:DCImage}(b).  This induced a significant amount of micromotion, but the single photon counting method is relatively (though not completely) insensitive to micromotion. To quantify the affect, $DE_{SO}$ dropped by 7.6\% when the same voltage was applied, compared to the compensated $DE_{SO}$.  Shifting the ion towards the optic did increase the overall detection efficiency to $0.236(5)\%$, compared to 0.023(2) $\%$ when compensated.  This was still significantly lower than the theoretical value of $DE_{DOE}^{Expected}= 0.51\%$ (which includes the fractional reduction due to micromotion).
		 
\begin{figure}[htp]
\includegraphics[width=\columnwidth]{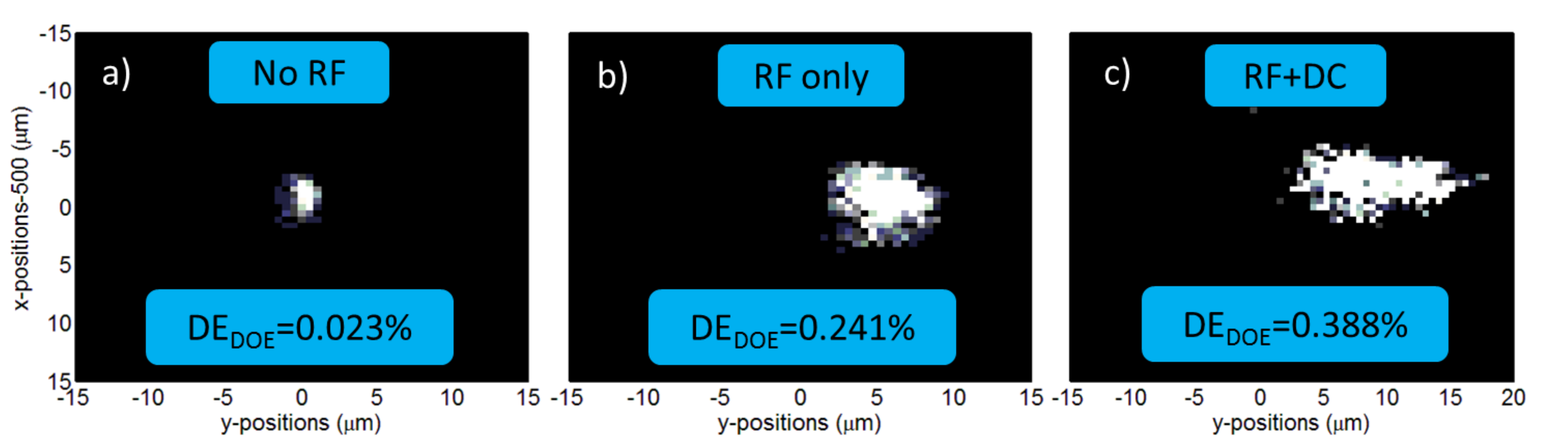}
\caption{\small(Color online)  (a) Image of a compensated ion (b) Image of a shifted ion due to applying 10 V$_{RF}$ to the center DC indicated by the star in Figure \ref{fig:trap}. (c) Ion after applying both an RF shift and a DC push of the ion.}
\label{fig:ImageRF}
\end{figure}  
		 
It was possible to improve $DE_{DOE}$ further by applying an additional RF voltage to one of the center DC electrodes (indicated in Figure \ref{fig:trap} with a $^*$). By reducing the amplitude of the RF and phase locking the signal applied to one center electrode, the RF null shifted toward the center of the DOE.  Figure \ref{fig:ImageRF} shows a series of three images and the corresponding $DE_{DOE}$ at each location.  Figure \ref{fig:ImageRF}(a) shows the image before applying RF to the center DC and figure \ref{fig:ImageRF}(b) shows the shift while applying 10 $V_{RF}$ of phase locked RF, resulting in an ion displacement of 8 $\mu$m and a $DE_{DOE}$ of 0.241(5)\%.  Figure \ref{fig:ImageRF}(c) shows the ion displaced by 10-12 $\mu$m when both RF and DC fields are used.  In this case $DE_{DOE}=$ 0.388(6)\%, less than the expected $DE$ of $0.51$\% due to the remaining imperfect overlap of the ion with the DOE focus.
		
\section{Conclusion}\label{sec:conclusion}

We demonstrated an integrated array of DOE's with a surface electrode ion trap, coupling fluorescence from a trapped Ca$^{+}$ ion into a multi-mode fiber under ultra high vacuum conditions and successfully transmitting the light out of the vacuum chamber to a PMT. Stray electric fields were easily compensated over the optic and remained stable over the course of weeks, and the DOE did not noticeably affect the motional heating rate of the ion.  Using a single photon counting technique we measured an overall detection efficiency for the DOE of 0.388\%.
	
\section{Acknowledgments}

We would like to thank Robert Boye for many helpful discussions, and M. G. Blain, Joel Wendt, Sally Samora, and the entire fabrication team for providing us with the trap and optics. We also thank Greg Brady for the optical design. This work was supported by the Intelligence Advanced Research Projects Activity (IARPA). Sandia National Laboratories is a multi-program laboratory managed and operated by Sandia Corporation, a wholly owned subsidiary of Lockheed Martin Corporation, for the US Department of Energy's National Nuclear Security Administration under contract DE-AC04-94AL85000.


\end{document}